\documentclass[aps,prx,floatfix,twocolumn]{revtex4-1}

\usepackage[english]{babel}

\makeatletter
\def\bbl@set@language#1{%
	\edef\languagename{%
		\ifnum\escapechar=\expandafter`\string#1\@empty
		\else\string#1\@empty\fi}%
	\@ifundefined{babel@language@alias@\languagename}{}{%
		\edef\languagename{\@nameuse{babel@language@alias@\languagename}}%
	}%
	\select@language{\languagename}%
	\expandafter\ifx\csname date\languagename\endcsname\relax\else
	\if@filesw
	\protected@write\@auxout{}{\string\select@language{\languagename}}%
	\bbl@for\bbl@tempa\BabelContentsFiles{%
		\addtocontents{\bbl@tempa}{\xstring\select@language{\languagename}}}%
	\bbl@usehooks{write}{}%
	\fi
	\fi}
\newcommand{\DeclareLanguageAlias}[2]{%
	\global\@namedef{babel@language@alias@#1}{#2}%
}
\makeatother

\DeclareLanguageAlias{en}{english}

\usepackage{graphicx}
\usepackage{epstopdf}
\usepackage{amsmath}
\usepackage{lineno}
\usepackage{comment}
\usepackage{graphicx, nicefrac}
\usepackage{float}
\usepackage[bookmarks=True,bookmarksopen=True]{hyperref} 
\usepackage{silence}
\usepackage[usenames,dvipsnames]{color}
\usepackage{colortbl}
\newcommand{\TV}[1]{} 
\addto\captionsenglish{}
\addto\captionsenglish{}

\newcommand{\RN}[1]{%
  \textup{\uppercase\expandafter{\romannumeral#1}}%
}

\hypersetup {
	unicode=false,          
	pdftoolbar=true,        
	pdfmenubar=true,        
	pdffitwindow=false,     
	pdfstartview={FitH},    
	pdftitle={High-resolution Energy Control of Ultracold Atom-ion Collisions},    
	pdfauthor={Ruti Ben-shlomi},     
	pdfcreator={Ruti Ben-shlomi},   
	colorlinks=true,       
	linkcolor=BlueViolet,          
	citecolor=BlueViolet,        
	filecolor=Black,      
	urlcolor=Black           
}

\graphicspath{{./figures/}}  

\usepackage[titletoc,title]{appendix}

\begin{document}
	\title{High-energy-resolution measurement of ultracold atom-ion collisional cross section}
	
	\author{Ruti Ben-shlomi$^1$}\email{ruti.ben-shlomi@weizmann.ac.il} 
	\author{Meirav Pinkas$^1$}
	\author{Ziv Meir$^1$}\altaffiliation{Present address: Department of Chemistry, University of Basel, Klingelbergstrasse 80, CH-4056 Basel, Switzerland}
	\author{Tomas Sikorsky$^1$}\altaffiliation{Present address: Atominstitut - E141, Technische Universität Wien, Stadionallee 2, 1020 Vienna, Austria}
	\author{Or Katz$^1$}
	\author{Nitzan Akerman$^1$}
	\author{Roee Ozeri$^1$}
	
	
	
	\affiliation{$^1$ Department of Physics of Complex Systems, Weizmann Institute of Science, Rehovot 7610001, Israel.}
	
\date{\today}
\begin{abstract}
{The cross section of a given process fundamentally quantifies the probability for that given process to occur. In the quantum regime of low energies, the cross section can vary strongly with collision energy due to quantum effects. Here, we report on a method to directly measure the atom-ion collisional cross section in the energy range of 0.2-12 mK$\cdot$ k$_B$, by shuttling ultracold atoms trapped in an optical-lattice across a radio-frequency trapped ion. In this method, the average number of atom-ion collisions per experiment is below one such that the energy resolution is not limited by the broad (power-law) steady-state atom-ion energy distribution. Here, we estimate that the energy resolution is below 200 $\mu$K$\cdot$k$_B$, limited by drifts in the ion's excess micromotion compensation and can be reduced to the 10's $\mu$K$\cdot$k$_B$ regime. This resolution is one order-of-magnitude better than previous experiments measuring cold atom-ion collisional cross section energy dependence. We used our method to measure the energy dependence of the inelastic collision cross sections of a non-adiabatic Electronic-Excitation-Exchange (EEE) and Spin-Orbit Change (SOC) processes. We found that in the measured energy range, the EEE and SOC cross sections statistically agree with the classical Langevin cross section. This method allows for measuring the cross sections of various inelastic processes and opens up possibilities to search for atom-ion quantum signatures such as shape-resonances.} 
\end{abstract}                                          
\maketitle

\section{Introduction}
Ultracold atom-ion systems have emerged in the last decade as a fast-growing field and have gained large interest due to their potential contribution to quantum chemistry \cite{Ratschbacher2012,Puri2017,Chang2013,Sik2018}, quantum computing \cite{Doerk2010,Gerritsma2012} and quantum simulation \cite{Joger2014,Secker2016} fields.

Collisions between atoms and ions are characterized by an attractive
long-range polarization-potential which scales as $-r^{-4}$ and leads to a semi-classical behavior over a wide range of collision energies \cite{Saito2017}. At very low energies, quantum phenomenon, such as Feshbach \cite{Idziaszek2009,Idziaszek2011,Tomza2015} and shape resonances \cite{Silva2015,Raab2009,Tacconi2011,Belyaev2012}, are predicted, similarly to the one observed in atom-atom \cite{Inouye1998} and atom-molecules \cite{Klein2017} collisions. Therefore, there is a considerable experimental effort for cooling atom-ion mixtures into the few partial-wave regime and measuring the energy dependence of the cross section for different collisions and reactions with high resolution.

Reaching the few partial-wave regime in atom-ion systems has been a significant challenge for the atom-ion community in the last couple of decades. The reason being that at steady-state, the collision energy between atoms and ions is neither fundamentally limited by the temperature to which both species are cooled, nor by the ion’s trap residual Excess-Micromotion (EMM) energy. Instead, this fundamental limit is set by the force that the atom exerts on the ion during a collision. This force is then amplified by the ion-trap oscillating fields \cite{Cetina2012,Meir2016,Pinkas2020,Feldker2020}.
This effect sets the lower bound of atom-ion steady-state interaction energy in these systems. 
Up until recently, this lower bound has been at least two orders of magnitude higher than the s-wave energy limit \cite{Schmid2018}.
Nevertheless, this fundamental energy limit is species dependent \cite{Cetina2012}, and favorable for  mixtures combined by light-atoms and heavy-ions such as $^6$Li-Yb$^+$. Only recently researchers have reached the s-wave regime for that system, with collisional energies of about 10 $\mu$K$\cdot$k$_B$ \cite{Feldker2020}. \\

In recent years, several experiments studied the rates and cross sections of inelastic atom-ion collisions as a function of collision energy. Several experiments reached the energy regime where quantum resonances should appear \cite{Hall2013, Hall2012, Haze2013,Saito2017, Dorfler2019}, but these have yet to be observed. In all previous studies, scanning the energy was accompanied by an  increase of the energy spread, and thereby compromising the energy resolution. In one method, the collision energies were varied by increasing the micromotion energy of the ions, which is associated with their motion in the oscillating rf, electric field \cite{Zipkes2010,Hall2013,Schmid2010,Hall2012,Schmid2012,Haze2013}. However, increasing excess micromotion broadened the ion energy spread into a power-law distribution in which the distribution spread was larger than the distribution peak \cite{Hall2012,Silva2015,Bell2009,Grier2009,Puri2019}. In a different experiment a magneto-optical trap of atoms was shuttled across a crystal of atomic \cite{Eberle2016} or molecular ions \cite{Dorfler2019}, using radiation-pressure forces, reaching an energy resolution in the mK regime. Another approach is shuttling the ion by modulating the voltage on the trap electrodes \cite{Puri2018}. By these methods the collision energy can be scanned between $\sim$10 mK to $\sim$1 K with a relative resolution of $\sim$10. In the method presented here, the inferred energy resolution of $\sim$200 $\mu$K$\cdot$k$_B$ is at-least one order of magnitude narrower. \\ 

Here we present a method for high energy-resolution control of atom-ion collision by shuttling a cloud of ultracold atoms, trapped in a one-dimensional optical lattice, across a single trapped ion while maintaining a narrow collision energy spread. The collision energy is scanned with high resolution by changing the frequency difference between the optical lattice beams. We avoid the limitations imposed by the steady-state atom-ion energy distribution by limiting the average number of Langevin collisions in each pass to be smaller than one. Thus, the broadness of the collision energy is determined by the ion's and atoms' energy distributions prior to the collision both of which are in the 10's $\mu$K$\cdot$k$_B$ regime. For that, this method has sufficient energy resolution to potentially allow for the observation of quantum signatures such as shape resonances.  \\

We demonstrated our method by measuring the energy dependence of the inelastic collisions cross sections of the Electronic-Excitation Exchange (EEE) and Spin-Orbit Change (SOC), channels that occur when a $^{88}$Sr$^+$ ion, optically excited to the 4d$^2$D$_{5/2}$ meta-stable state, collides with ground-state $^{87}$Rb atoms. These processes were shown \cite{Ruti2019} to occur through a non-adiabatic Landau-Zener crossing and their energy dependence was only theoretically discussed up to now \cite{Belyaev2012,Hall2011}, for the same collisional energy range. We measured the energy dependence of the inelastic collision cross section of the EEE channel and the SOC channels separately. We found that for collision energies ranging between 0.2$-$12 mK$\cdot$k$_B$, the cross-section for both channels follow the semi-classical Langevin $E^{-1/2}$ scaling with good statistical significance.

Finally, we discuss in this manuscript the effect of multiple collisions on the energy resolution of our method and also analyze possible deviations from the semi-classical Langevin scaling, in search of quantum resonances, by performing a maximum-likelihood estimation test. 

\section{EXPERIMENTAL SET-UP}
An illustration of our experimental setup is shown in Fig.~\ref{ESC}. Our hybrid atom-ion system is described in detail in a review article \cite{Meir2018}. The setup consists of two separate vacuum chambers. At the top chamber $\sim$5$\cdot$10$^7$ cold Rb atoms are trapped in a magneto-optical-trap and then loaded into a CO$_2$ trap to evaporatively cool the atoms to the $\sim$5 $\mu$K$\cdot$k$_B$ temperature range. At the end of the evaporation, $\sim$50,000 atoms remain in the CO$_2$ trap and are adiabatically loaded into a 1D optical lattice. The lattice consists of two counter-propagating YAG laser beams ($
\lambda=1064$ nm, $P=1.5$ Watt for each beam), which are collimated, vertically orientated and have a Gaussian profile \cite{Schmid2006}. The beams are characterized by a waist of $\sim$220 $\mu$m and a Rayleigh range of $z_R = \pi w^{2}_{0}/\lambda=143$ mm, comparable to the transport distance to the bottom chamber of 248 mm. The strong confinement of the atoms in the optical lattice sites in the transport direction prevents the loss of atoms due to gravity.  
We shuttle the atoms to the bottom chamber by changing the relative frequency between the two lattice beams (further details in the next section). During the transport, a $^{88}$Sr$^+$ ion is held in a linear segmented rf Paul trap, optically pumped to a specific Zeeman state in the electronic ground state, 5s$^2$S$_{1/2}(m=-1/2)$, followed by ground-state cooling on all three motional mode to $\bar{n}<0.1$. We repeatedly compensate the EMM of the ion every roughly half an hour during the experiment to avoid EMM drifts. 

A thorough analysis of the EMM in our system yields that the sum of all EMM contributions is $\sim$30 $ \mu$K$\cdot$k$_B$ \cite{Meir2018}. This number can be used to estimate the lower bound for the energy resolution of this method in our system. Here, however, due to drifts in the micromotion compensation during the experiment, we set an upper limit for the EMM in our system to be $\sim$200 $ \mu$K$\cdot$k$_B$ which sets the limit for the resolution in the experiments presented here.

\section{COLLISION VELOCITY CONTROL}
We set the relative velocity of the atoms  compared to the stationary ion by controlling the relative frequency of the lattice beams. The atoms’ velocity is directly proportional to the instantaneous frequency difference between the beams, $\Delta f(t)$, and equal to $v(t)=\frac{\lambda\Delta f(t)}{2}$ in the lab-frame where $\lambda=$1064 nm is the laser wavelength.
The linear velocity of the atoms in the lattice is much higher than the thermal velocity of the atoms or the ion. Therefore, the atom-ion collision energy is set by the velocity of the lattice.
In order to transport the cloud of atoms across the trapped-ion in a well defined collision energy, $E_{coll}=\frac{1}{2}m_{Rb}v_{lattice}^2$, in the lab-frame, the frequency difference between the laser beams should satisfy,
\begin{equation}
    \label{coll_freq}
    \Delta f(t)=2\sqrt{(2E_{coll}/m_{Rb})}/\lambda.
\end{equation}
Here, $m_{Rb}$ is the mass of Rb atom.
\begin{figure*}[t!]
\begin{center}
\centering{\includegraphics[width=\textwidth]{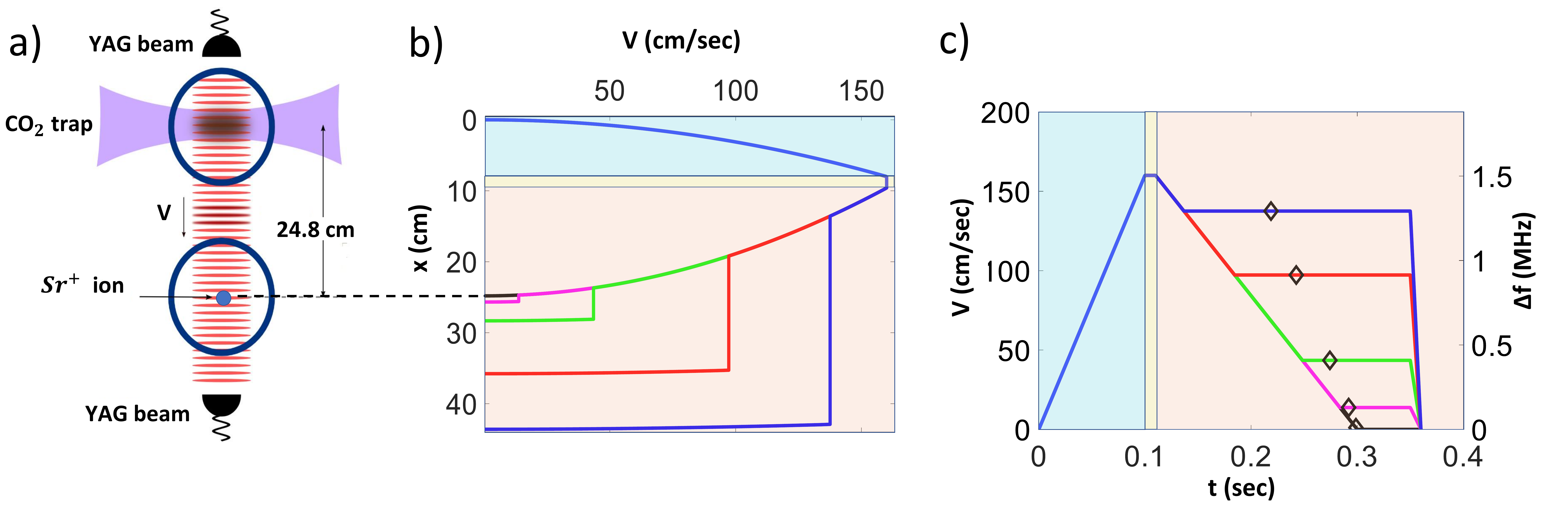}}
\caption{An illustration of the atom-ion system and the velocity profiles of the transport. a) The experiment setup. Rb atoms are held in a CO$_2$ dipole trap, loaded into the 1D, optical lattice made of two counter-propagating beams. The atoms occupy $\sim$40 lattice sites. The distance between the loading position of the atoms in the upper chamber, and the ion in the lower chamber is 248 cm, as indicated by the dotted black curve. The figure is not to scale. b) The atoms' position as function of their velocity, for five different collision energies: 0, 0.1, 1, 5, 10 $k_B\cdot$mK (from black to blue). The background represents the different stages of transport: acceleration (pale blue), movement in a constant velocity (yellow) and deceleration to the desired collision velocity (pink). The atoms continue to be transported by the lattice at a constant velocity, even after colliding the ion. c) The atoms velocity profile. The right y-axis $\Delta f$, is the corresponding frequency difference. Each line is the difference between the same color lines (solid and dotted-dashed) in Fig.~\ref{freq}. The diamond marks indicate the times for which the atoms collide the ion for each velocity profile.} 
\label{ESC}
\end{center}
\end{figure*}
The two lattice beams pass through separate acousto-optic modulators (AOMs) in a double-pass configuration, to control their frequency and intensity. After the AOMs, the beams are coupled to fibers, one entering from above the atoms' chamber and the other from below the ion's chamber, as illustrated in Fig.~\ref{ESC}a. When varying the frequency upon the AOMs, different diffraction angles cause a change in the intensity of the beams. Therefore, we actively stabilize the intensity level of each beam, maintaining constant intensity throughout the entire experiment. Each lattice beam is connected to a separate frequency channel of a function-generator capable of generating a trapezoidal sweep of the frequency independently in each channel. The two trapezoidal sweeps combine to generate the relative frequency profile, $\Delta f(t)$, shown in Fig. 1c. (see Appendix). 

To bring the atoms to the desired velocity when colliding with the ion, we control the frequencies of both lattice beams. 
We design the frequency profile such that the atoms always accelerate to the same maximal velocity, and then decelerate to the desirable collision velocity, as can be seen in Fig.~\ref{ESC}b and c. 
The atoms reach a maximal velocity of $v$=160 cm/sec after 0.1 seconds of acceleration. Then, the atoms are held at a constant velocity for 0.01 seconds, after which the atoms are decelerated to the desired velocity.
For this velocity regime, the transport itself involves negligible atoms loss, where higher velocities introduce losses. 

The travelled distance of the atoms as function of their instantaneous velocity is shown in Fig.~\ref{ESC}b for different transport profiles. After the first 0.11 seconds, the atoms are transported 9.6 cm. From this point, the atoms start to decelerate until they arrive to the desired velocity.
For each collision energy, the atoms cease to decelerate at a different position relative to the ion. The atoms continue to move at a constant velocity until they pass the position of the ion.

\section{Measuring inelastic collisional cross sections}

The rate at which a given inelastic collisions occur is given by 
\begin{equation}
    \Gamma_{inelastic}=n_{atoms}\sigma(E_{coll})v_{coll}
\end{equation}
where $n_{atoms}$ is the atomic density, $\sigma(E_{coll})$ is the inelastic collisional cross section which is energy dependent, and  $v_{coll}$ is the relative atom-ion velocity in the center-of-mass frame.

In the proposed scheme, the atomic density in the position of the ion is time-dependent due to the relative motion of the atoms in the lattice with respect to the stationary ion. The mean number of collisions per pass is given by
\begin{equation}
    N=\sigma(E_{coll})v_{coll}\int_{-\infty}^{+\infty}n_{atoms}(t)dt.
\end{equation}
Since the atoms moves at a constant velocity, $v_{lattice}$, the integration can be taken over a spatial dimension, in the moving direction of the lattice,
\begin{equation}
    N=\sigma(E_{coll})\frac{v_{coll}}{v_{lattice}}\int_{-\infty}^{+\infty}n_{atoms}(x)dx.
\end{equation}
Here, the temperatures of the ion and atoms are negligible relative to the velocity of the atoms in the lattice, and since the ion is stationary, this collision velocity is equal to the lattice velocity $v_{coll}=v_{lattice}$, in the lab frame.
Then, the number of collisions per pass is
\begin{equation}
    N=\sigma(E_{coll})\int_{-\infty}^{+\infty}n_{atoms}(x)dx.
    \label{N_coll}
\end{equation}
Therefore, the number of events we measure is directly proportional to the collisional cross section, through the density of the atoms in the lattice, integrated along the vertical direction of motion.

Assuming the length of the atomic cloud is finite, denoting it by $L_{Rb}$, we can rewrite the number of event as:
\begin{equation}
    N=n_{eff}L_{Rb}\sigma(E_{coll}),
    \label{crosssection}
\end{equation}
where we define an effective density as: 
\begin{equation}
    n_{eff}\equiv\frac{1}{L_{Rb}}\int_{-\infty}^{+\infty}n_{atoms}(x)dx.
\end{equation}

In the semi-classical regime, the total cross section for hard-sphere collisions between an ion and an atom is given by the Langevin cross section \cite{Langevin1905}:
\begin{equation}
    \sigma_L=\pi \sqrt{\frac{2C_4}{E_{coll}}},
\end{equation}
Where $C_4=\alpha e^2/(4\pi\epsilon_0)^2$, with $\alpha$, $e$ and $\epsilon_0$ the atoms polarizability, electronic charge and the vacuum permittivity, respectively.

Thus, the mean number of Langevin collisions per pass is
\begin{equation}
    N_L=\pi n_{eff} L_{Rb} \sqrt{\frac{2C_4}{E_{coll}}}.
\end{equation}
In the semi-classical regime, inelastic processes are proportional to the Langevin cross section and therefore scale as $\sim E_{coll}^{-1/2}$.

\section{ENERGY DEPENDENCE OF NON-ADIABATIC QUENCH OF META-STABLE EXCITED STATES }
To demonstrate our method, we measured the energy dependence of a non-adiabatic quench of a meta-stable electronically excited level of the ion during a collision with a ground-state atom . In previous work \cite{Ruti2019}, we found that the excited long-lived 4d$^2D_{5/2}$ and 4d$^2D_{3/2}$ states of the $^{88}$Sr$^+$ ion, quench after roughly three Langevin collisions with ground-state $^{87}$Rb atoms, and that the excitation energy is transformed into kinetic energy of the colliding particles. 

In Ref.~\cite{Ruti2019} we identified two types of collisional quenching. One is the EEE where the ion relaxes to the ground S state and the atom is excited to the P state followed by energy release of $\sim$ 3000 K$\cdot$k$_B$. The second is SOC where the ion relaxes from the higher fine-structure D$_{5/2}$ level to the lower D$_{3/2}$ level releasing $\sim$ 400 K$\cdot$k$_B$ into kinetic energy. These processes were theoretically understood to occur through Landau-Zener avoided-crossings between the different molecular potential curves. \\

Here we measured the dependency of these inelastic cross sections on the collision energy. As described above, a single Sr$^+$ ion, cooled to its ground-state in all three motional modes and with a residual EMM bounded by $\sim$200 $\mu$K$\cdot$k$_{B}$, was prepared in the 4d$^2$D$_{5/2}(m=-5/2)$, lower Zeeman state. Here, we report a higher bound on the EMM value than what was reported in Ref. \cite{Meir2018} since it was compensated less often due to long interrogation times. Meanwhile, a cloud of un-polarized atoms was loaded into the optical lattice and shuttled to the lower chamber while scanning 119 energy points, from 0.2 to 12 mK in the lab-frame with energy steps of 100 $\mu$K$\cdot$K$_{B}$. The average number of Langevin collisions per sweep was tuned to be 0.09 in the lowest energy point.

\begin{figure}
\begin{center}
\centering{\includegraphics[width=\columnwidth]{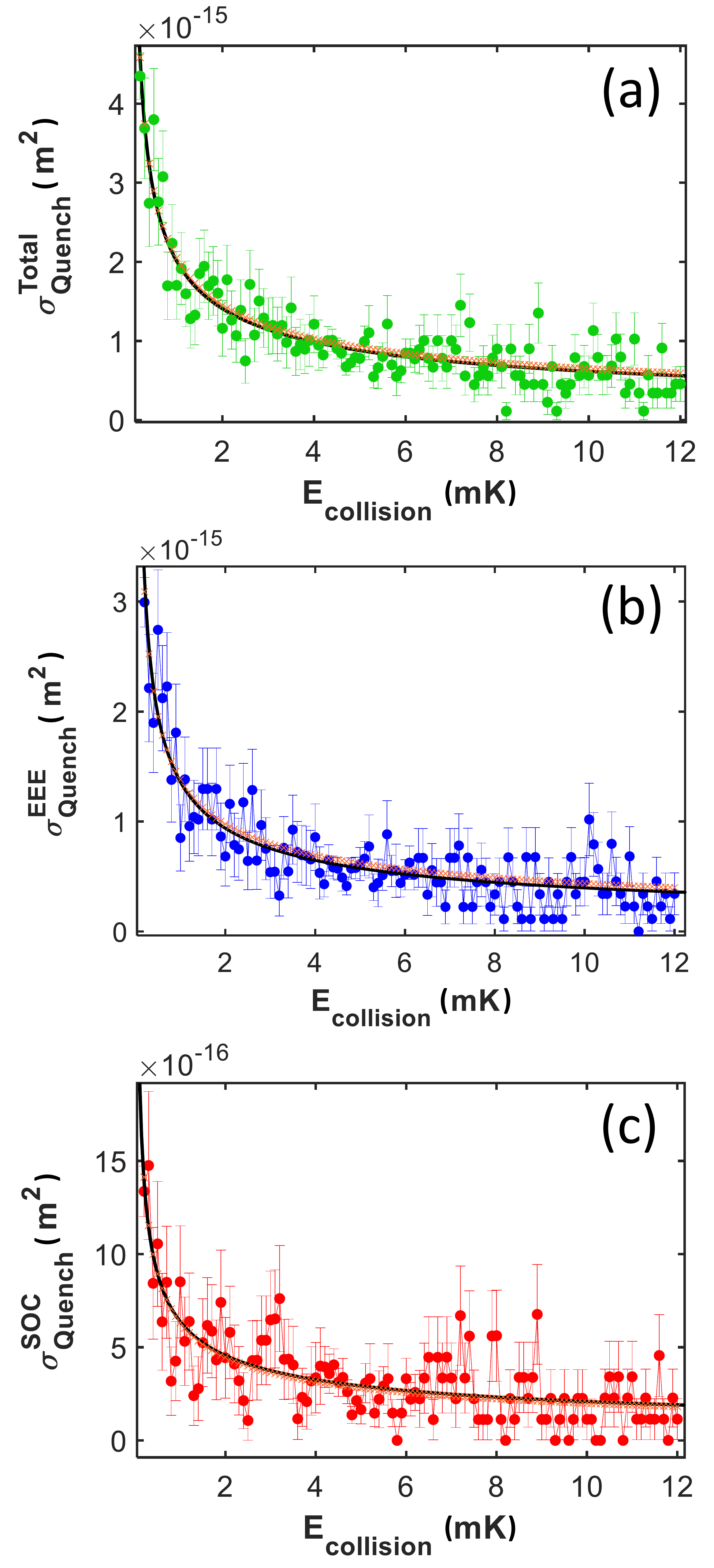}}
\caption{Quench cross section as a function of the collision energy. (a) The total quench probability (green) and for the (b) EEE (blue) and (c) SOC (red) channels, separately. Error-bars are 1$\sigma$ standard deviation of a binomial distribution and also includes systematic noises as explained in the body of the text. Smaller error-bars indicate areas over which we performed more repetitions. Black curve is an exponential fit to A$\cdot $E$^{\alpha}$, where the exponents given by the fits are $\alpha$=-0.51(3), -0.53(4) and -0.48(6), respectively. The orange crossed lines are fits to the Langevin cross section multiplied by a pre-factor $\eta$ as a free parameter, given by Eq.~\ref{langsigma}.}   	
\label{EnergyQuench}
\end{center}
\end{figure}

After the atoms passed through the ion, we performed a single-shot Doppler thermometry \cite{singleshot} on the ion to detect the quenched (hotter than $\sim$10's K$\cdot$k$_B$) events from non-quenched events. Due to the large energy separation between the SOC (400 K$\cdot$k$_B$) and EEE (3000 K$\cdot$k$_B$) energy release, these events are easily separated in the single-shot thermometry  \cite{Ruti2019}. As a control experiment we tested whether quench events are detected in the absence of atoms in the optical lattice. Since no hot events were observed in the absence of atoms, we concluded that our measurement had no false positive detection of quench events. To avoid accumulative systematic noises, we scanned the energy of the collision in a randomized manner performing a single experiment for each energy value and only then repeating the experiment to accumulate the signal. 
The quench data presented in Fig.~\ref{EnergyQuench}a was derived from 300,000 repetitions in which 3100 quench events were identified. With a repetition time ranging between 1 to 10 sec, depending on the quench channel and the Doppler-recooling time, this data was integrated over weeks.
\\

In Fig.~\ref{EnergyQuench}, we present our measured results. We plot the quench cross section as a function of the relative collisional energy through the relation of Eq.~\ref{crosssection}:
\begin{equation}
    \sigma_{Quench}(E_{coll})=\frac{N_{Quench}}{n_{eff} L_{Rb}}.
\end{equation}
In this experiment the typical, effective, atomic density is $n_{eff}$=4.4$\cdot10^{17}$ m$^{-3}$. The collisional velocity ranging from $v_{coll}$=19.4 cm/sec up to 150 cm/sec, corresponds to collision energies of 0.2 to 12 mK$\cdot$K$_{B}$, respectively. The size of the cloud in the transport direction is $L_{Rb}=$20 $\mu$m, which occupies $\sim$40 lattice sites.  
Here, different sets of data where taken with different cloud densities, adding systematic noise, biasing the overall data in the vertical direction upon the graph.
This systematic noise is estimated to be $\Delta (n_{eff}\cdot L_{Rb})\sim15\%$. The statistical noise for comparison, varies between 5$\%$ to 30$\%$, depending on the number of repetitions.

The data presented in green in Fig.~\ref{EnergyQuench}a, contains all quench events summing over both channels. In Fig.~\ref{EnergyQuench}b and Fig.~\ref{EnergyQuench}c we show the energy dependent collisional cross section for EEE (blue) and SOC (red) channels, respectively. The black curves are fits to a power-law, A$\cdot$E$^{\alpha}$. The fitted power-law, $\alpha$, agrees well with the Langevin scaling of E$^{-1/2}$ (see Fig.~\ref{EnergyQuench} caption). Quenching from the metastable D-state happens when the atom and ion reach to very short inter-nuclear distances and once overcoming the centrifugal barrier. Therefore these type of collisions are Langevin collisions, but happen with lower probabilities. Here we compare the cross sections through:
\begin{equation}
    \sigma_{Quench}(E_{coll})=\eta\cdot\sigma_{L}(E_{coll})
    \label{langsigma}
\end{equation}
By fitting the data to Eq.~\ref{langsigma} with $\eta$ as a free parameter, we find that 
$\sigma_{Quench}(E_{coll})$ is proportional but smaller than the Langevin cross section by: $\eta$=0.52(6), 0.35(5), 0.16(3) for the green, blue and red data, respectively. While this total cross section is slightly higher than the one reported in previous studies \cite{Ruti2019} (0.52(6) compare to 0.38(5)), the ratio between the two channels $\sigma^{SOC}_{Quench}$/$\sigma^{EEE}_{Quench}$ agrees within the statistical error (0.48(11) compare to 0.39(5) in the previous measurements). 

\section{The effect of multiple collisions}
In this experiment the atoms are much colder than the ion and therefore the energy resolution of our measurement is mainly limited by the energy uncertainty of the latter. Since the ion is cooled to the ground-state of all it's secular motional modes, the initial residual energy, prior to the collision, is mainly due to the residual EMM.

However, after a collision the energy of the ion can be changed due to coupling of the EMM to the ion's external degrees of freedom \cite{Zipkes2011} or due to exchange of kinetic energy between the atom and the ion \cite{DeVoe2009,Chen2014}. Both these effects depend on the position and phase of the ion in the rf trap and lead to a power-law energy distribution \cite{Rouse2017}.
Thus, in determining the energy spread of the ion before a reaction occurs, we have to take into account the possibility of ion heating due to previous, elastic, Langevin collisions.

\begin{figure}
    \centering
    \includegraphics[width=\columnwidth]{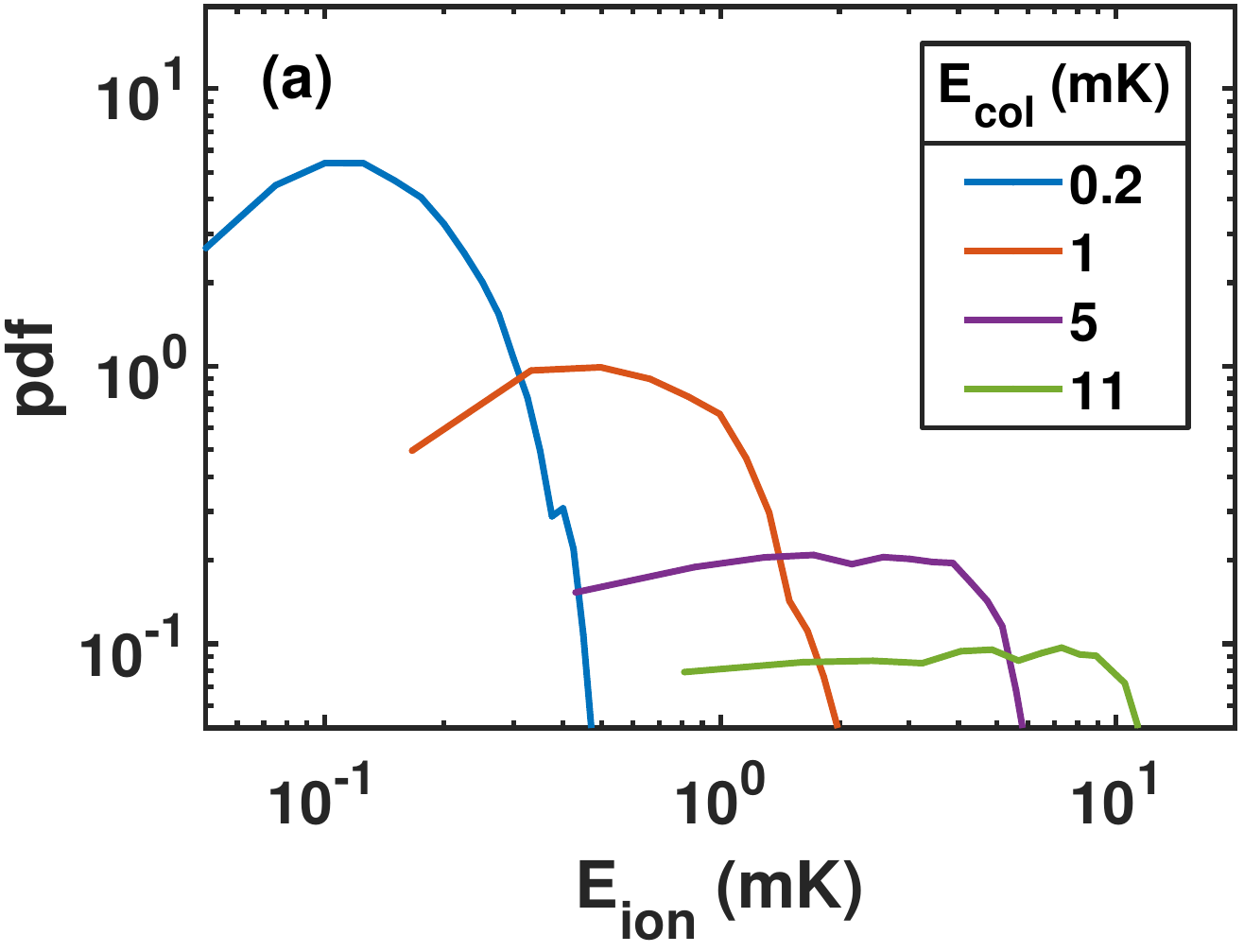}
    \caption{(a) The distribution of ion's energy, $E_{ion}$, after a single collision with an atom in the moving optical lattice, for different lattice velocities, $E_{col}$. The distributions are calculated by a molecular dynamics simulation that takes into account the Paul trap potential and the EMM of the ion.}
    \label{ion_energy_dist}
\end{figure}

In order to find the ion's energy distribution after a certain average number of collisions, we performed a molecular dynamics simulation which takes into account the residual EMM of the ion and the lattice velocity, as described in Ref. \cite{Zipkes2011,Meir2018}. In Fig.~\ref{ion_energy_dist}, the energy distribution of the ion after a single collision is shown for different velocities of the lattice. As can be seen, following a single collision, the ion is heated up to the energy of the atoms in the lattice, with a wide energy distribution. As a result, if the measured inelastic process (for example, a quench) does not occur in the first collision, the energy of that collision is no longer defined by the velocity of the lattice and has a wide distribution.

The probability for multiple Langevin collision events can be reduced by lowering the density of atoms that are loaded into the lattice dipole trap. However, this leads to longer integration times. As an example, in the data of Fig.~ \ref{EnergyQuench} the probability for at least one Langevin collision per-pass was approximately 0.09, for low collisional energies. At such a low mean number of collision, the probability for observing a quench event that occurred after the first collision is $\sim9\cdot10^{-4}$, and hence the signal is not effected by heating due to multiple elastic collisions. However, this measurement lasted for several weeks. 

\section{A Search for quantum resonances using Maximum Likelihood Estimation}
A hallmark of quantum scattering in the low energy regime is the appearance of scattering resonances. Such resonances occur, for example, when the collision energy in the center-of-mass frame resonates with the energy of a quasi-bound molecular state by the centrifugal barrier of one of the partial waves involved. These shape-resonances are anticipated to occur in atom-ion collisions even in the mK energy range \cite{Silva2015,Belyaev2012,Tacconi2011}. In order to search for such resonances we performed a likelihood-ratio test, differentiating between resonance and no-resonance hypotheses, and calculated their statistical significance. \\

At each collision energy, $E_i$, the number of observed quench events is a random variable which follows a binomial distribution. The log-likelihood function for observing $k_i$ quench events out of $N$ repetitions is, up to a constant factor,
\begin{equation}
    \log\mathcal{L}(p_i|k_i,N_i)=k_i\log p_i+(N-k_i)\log(1-p_i),
\end{equation}
where $p_i$ is the probability for observing a quench event in a single experiment. 

The total log-likelihood for observing $\textbf{k}=\{k_i\}$ quench events in all energy points is the sum over the log-likelihood function in each point,
\begin{equation}
    \log \mathcal{L}(\textbf{p}|\textbf{k},\textbf{N})=\sum_i\log \mathcal{L}(p_i|k_i,N_i).
\end{equation}

We want to estimate the probability that the data we measured is the result of a local peak at some energy point. The null hypothesis, $H_0$, assumes that the measured data has as a power-law $p_i(E_i)=CE^{-\alpha}$ behavior, whereas the alternative hypothesis includes a Gaussian resonance at energy $E_0$ with a width of $\sigma_{g}$ and magnitude $A$.
\begin{equation}
    p_i=CE^{-\alpha}+Ae^{\frac{(E_i-E_0)^2}{\sigma_{g}^2}}.
\end{equation}

We estimated the free parameters ($C$, $\alpha$, $A$, $\sigma$ and $E_0$) as the parameters that maximize the log-likelihood function for our measured data. Using these parameters, we calculated the observed likelihood-ratio between the null hypothesis and the alternative hypothesis

\begin{equation}
    \label{eq:likelihood-ratio}
\log \lambda_{obs}=\max \log \mathcal{L}_0- \max \log \mathcal{L}_1.
\end{equation}

In order to make the maximization process of the alternative hypothesis more robust, we found the maximum likelihood for a resonance separately for each energy point, and then identified the energy point that yielded the maximal likelihood as a suspect for resonance.

We used the likelihood-ratio of the measurement to estimate the statistical significance of the alternative hypothesis over the null hypothesis. To this end, we calculated the p-value: the probability of observing a likelihood-ratio that is higher than the one we measured under the null hypothesis. A small p-value indicates that it is less likely that our measured data was generated by the null hypothesis and the resonance hypothesis is favorable. The p-value can be related to the number of standard deviations, $N_\sigma$, of the observed data from the null hypothesis \cite{Demortier2007}
\begin{equation}
    p=1-\mbox{erf}(\frac{N_\sigma}{\sqrt{2}}),
\end{equation}
where $\mbox{erf(x)}$ is the standard error-function.

In order to find the p-value of our measurement, we simulated $1000$ experiments ($3000$ for the SOC experiment), each with the same number of repetitions we had in the real experiment, under the null hypothesis. For each one of these simulations, we repeated the analysis above in order to find the likelihood-ratio. From the simulated likelihood-ratio distribution we found the fraction of experiments that yielded a higher value than our observed likelihood-ratio, which gives the p-value of the measurement.

\begin{figure}
    \centering
    \begin{minipage}{0.4\textwidth}
    \includegraphics[width=1\linewidth]{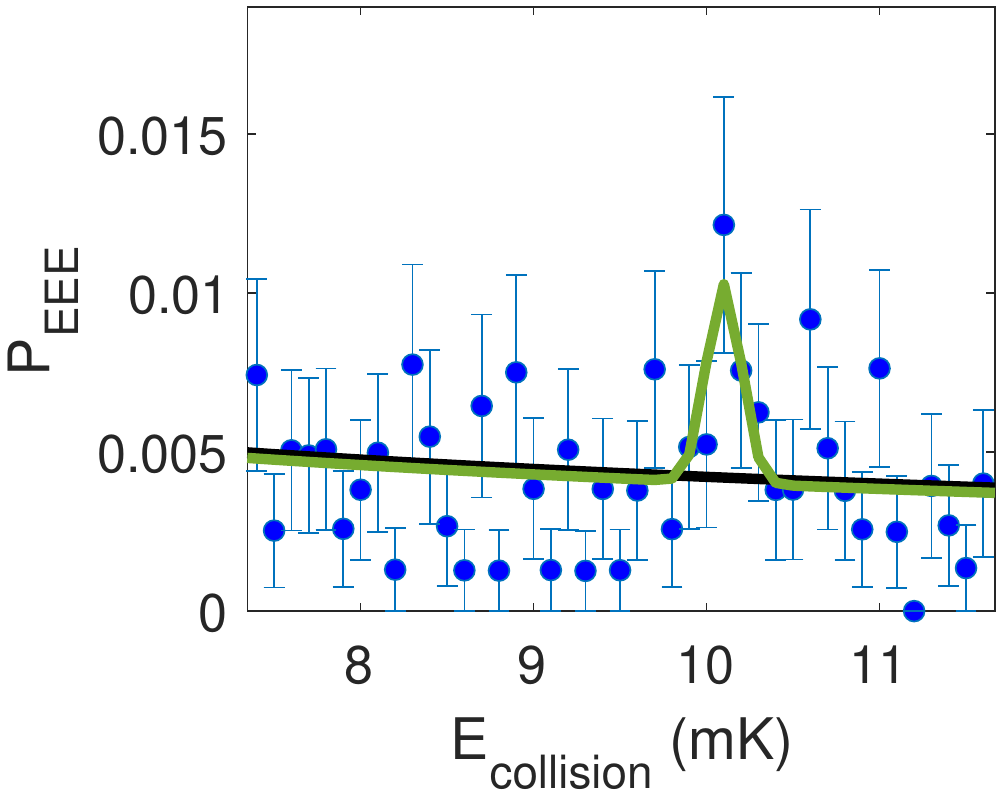}
    \end{minipage}
        \begin{minipage}{0.4\textwidth}
    \includegraphics[width=1\linewidth]{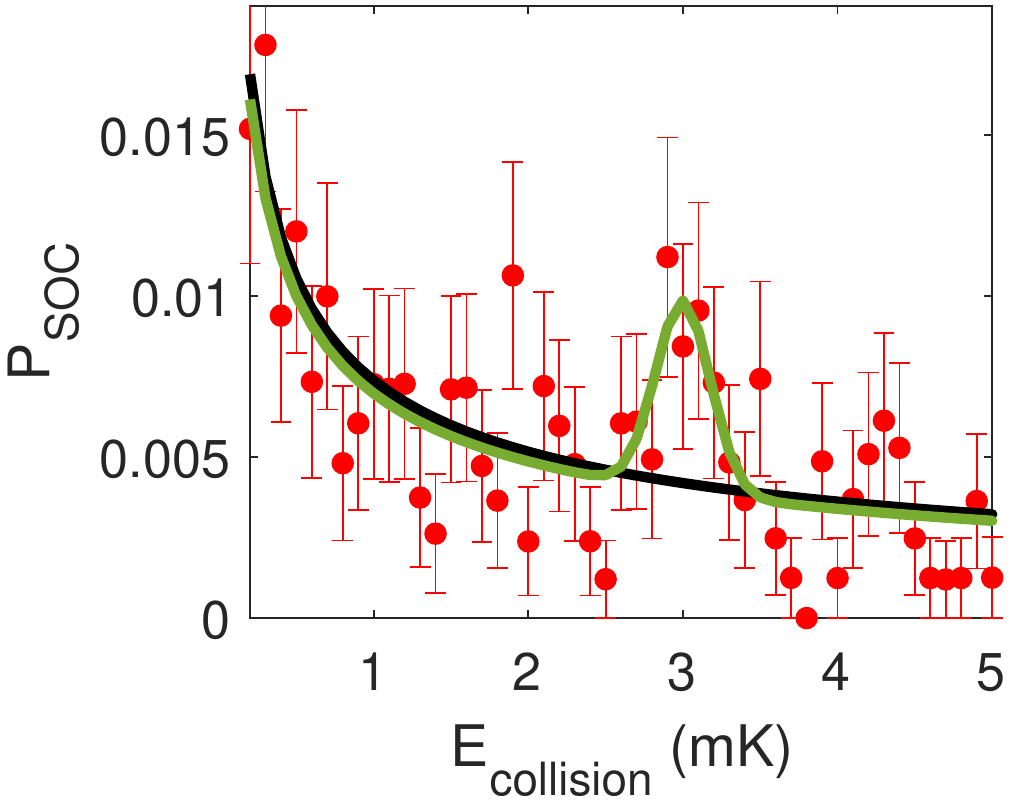}
    \end{minipage}
    \caption{Maximum likelihood estimation for a semi-classical cross section for the un-normalized, EEE channel (up) and SOC channel (down) for with and without a Gaussian resonance, green and black respectively. P-values are 0.091 and 0.0088, , which corresponds to 1.7$\sigma$ and 2.6$\sigma$, for the EEE and SOC channels respectively.}
    \label{fig:res_MLE}
\end{figure}

Analysing the EEE events, we observed a weak resonance at 10.3 mK with a likelihood ratio of 4.6 and a p-value of 0.091, equivalent to 1.7$\sigma$, see Fig.~\ref{fig:res_MLE}a. The analysis of the SOC events (Fig.~\ref{fig:res_MLE}b), indicated a peak around 3 mK with likelihood-ratio of 7.9. The p-value in this case is 0.0088, equivalent to 2.6$\sigma$, which is marginally significant.

Longer integration and improved statistics around suspected energies will help determine whether there is a resonance behavior or not. However, longer integration can suffer from systematic drifts that will wash-out the effect of a resonance. A further investigation of this effect with higher statistics is needed, with an improved repetition rate of the experiment to avoid drifts.
\section{conclusions}
In this work we present a method for controlling atom-ion collisional energy in the ultracold regime, with a one order of magnitude improved energy resolution as compared to previous methods, by optically shuttling the atoms across a single trapped-ion. The energy resolution of the reactions we study is maintained high by limiting the number of atom-ion collisions in each repetition of the experiment to be below one, and was thus limited only by EMM compensation to below 200 $\mu$K$\cdot$k$_B$ in this experiment with the potential to reach the 10's $\mu$K$\cdot$k$_B$ level with better control over the EMM over long periods of time. 

As a demonstration of our method, we used it to measure the energy dependence of the collisional quench processes of the ion from an optically excited meta-stable state. We found that the cross section for these processes follow the semi-classical Langevin prediction. Finally, we identified suspect energies for the possible location of a quantum resonance. Further experimental investigation is necessary to determine whether a resonance is actually present. 
Our method is generic and can be used for different species and for the study of different atom-ion reactions. With sufficient control of experimental parameters it can be used to measure atom-ion quantum scattering effects in the low partial-wave regime.

\section{Acknowledgments}
This work was supported by the Israeli Science Foundation and the Israeli ministry of Science Technology and Space. 

\section*{Appendix}
\subsection{Frequency profiles}
\begin{figure}
\centering{\includegraphics[width=\columnwidth]{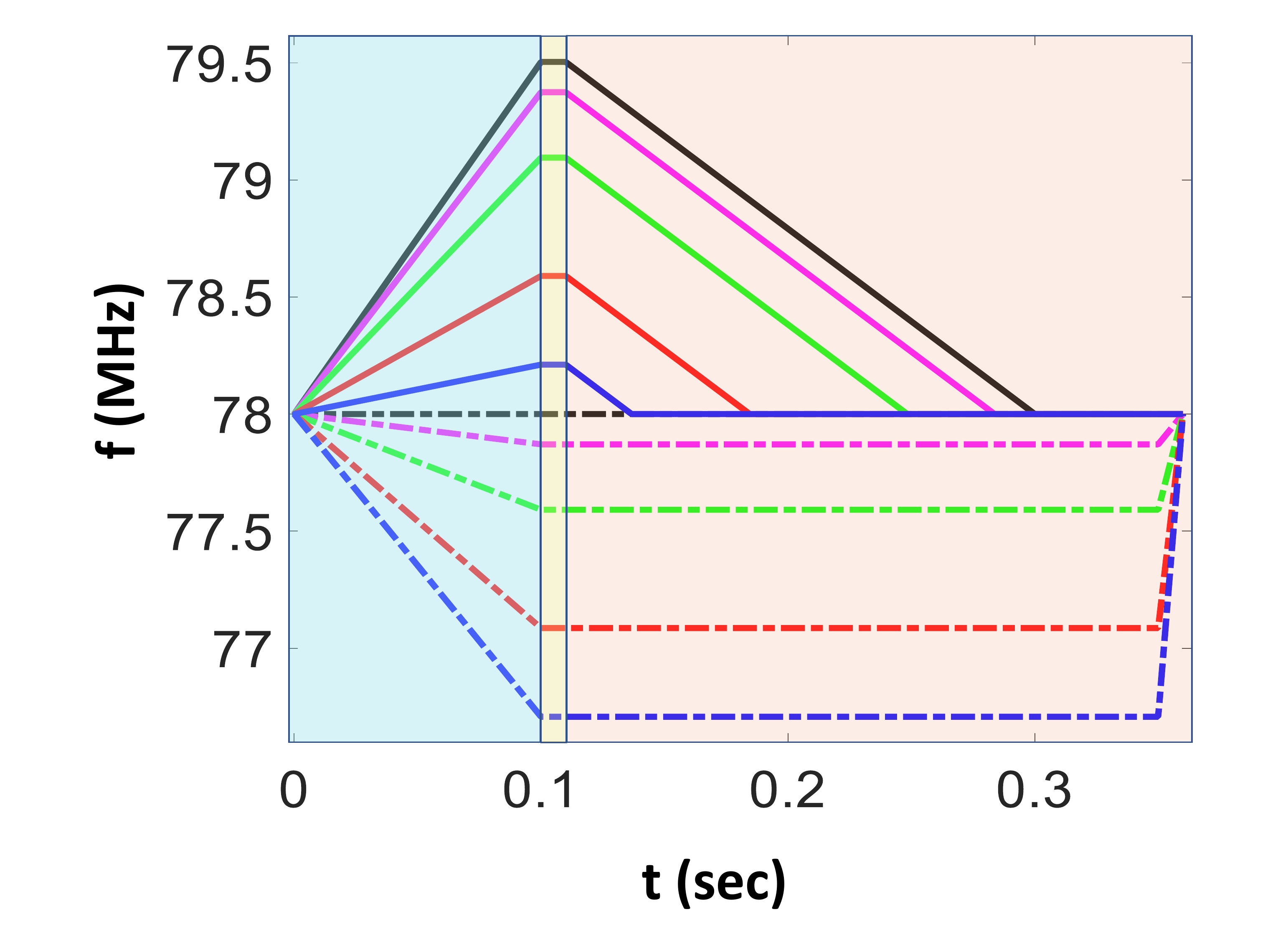}}
\caption{The frequency profiles of the two AOMs controlling the frequency of the  lattice laser beams. The solid (dotted-dashed) curve refers to the frequency set by the function-generator of the upper (lower) lattice beam.}   	
\label{freq}
\end{figure}

The trapezoidal frequency profile is characterized for each lattice beam by a start frequency $f_{s}$ and an intermediate frequency $f_{i}$ and three different stages: rise time in which the frequency is linearly increased from $f_{s}$ to $f_{i}$; hold time in which the frequency is kept constant at $f_{i}$; and return time in which the frequency is linearly decreased from $f_{i}$ to $f_{s}$ again. 
In order to move the atoms from the upper chamber to the ion in the lower chamber such that they will stop exactly on the ion, we use a trapezoidal frequency profile on one beam only, as illustrated by the black solid line in Fig.~\ref{freq}, while keeping the second beam at a constant frequency (black dotted-dashed line). We accelerate the atoms downwards for 0.1 seconds to a velocity of 83 cm/sec, then we keep them during 0.01 seconds under constant velocity, and then decelerate them in 0.2 seconds back to rest. The start frequency of both channels fed into both AOMs is set to $f_s$= 78 MHz. The intermidiate frequencies of the two AOMs are defined as:
$f^1_{i}=f_s+f_0-f_f$, $f^2_{i}=f_s-f_f$, 
where $f_0$=1503.9 kHz is the offset frequency corresponding to the maximal velocity of the atoms and $f_f$ is the final given frequency reproducing the desired collisional energy:
\begin{equation}
f_f=\Delta f(t)/2=\sqrt{(2E_{coll}/m_{Rb})}/\lambda,
\end{equation}
where factor of two comes due to the double-pass AOMs configuration. Note that for the second AOM, the hold time is longer than the time it takes the atoms to reach the ion in order to reach the correct final velocity.


\begin{references}	

\bibitem{Ratschbacher2012} L. Ratschbacher, C. Zipkes, C. Sias, and M. Köhl, \href{https://www.nature.com/articles/nphys2373}{Nat. Phys. \textbf{8}, 649–652 (2012)}.

\bibitem{Puri2017}  P. Puri, M. Mills, C. Schneider, I. Simbotin, J. A. Montgomery
Jr., R. Côté, A. G. Suits, E. R. Hudson,  \href{https://science.sciencemag.org/content/357/6358/1370.abstract?casa_token=_4AjTydJr_kAAAAA:JQloJiZw8He7hmwP6H2WYp_HmPq3R2L3tgxsW1K_0ERLJ1XbC-2Xa1xCzaCb0Wpkzsb11VGl097OUmU} {Science
\textbf{357}, 6358 (2017)}.

\bibitem{Chang2013}  Y. P. Chang, K. Długołęcki, J. Küpper, D. Rösch, D. Wild, S. Willitsch, \href{https://science.sciencemag.org/content/342/6154/98/tab-article-info} {Science
\textbf{342}, 6154 (2013)}.

\bibitem{Sik2018}  Y. P. Chang, K. Długołęcki, J. Küpper, D. Rösch, D. Wild, S. Willitsch, \href{https://www.nature.com/articles/s41467-018-03373-y} {Nat. comm
\textbf{9}, 920 (2018)}.

\bibitem{Doerk2010}  T. Sikorsky, Z. Meir, R. Ben-shlomi, N. Akerman and R. Ozeri  \href{https://journals.aps.org/pra/abstract/10.1103/PhysRevA.81.012708} { Phys. Rev. A \textbf{81}, 012708 (2010)}.

\bibitem{Gerritsma2012}  R. Gerritsma, A. Negretti, H. Doerk, Z. Idziaszek, T. Calarco,
and F. Schmidt-Kaler, \href{https://journals.aps.org/prl/abstract/10.1103/PhysRevLett.109.080402} {Phys. Rev. Lett. \textbf{109}, 080402 (2012)}.

\bibitem{Joger2014}  J. Joger, A. Negretti, and R. Gerritsma, \href{https://journals.aps.org/pra/abstract/10.1103/PhysRevA.89.063621} {Phys. Rev. A \textbf{89},
063621 (2014)}.

\bibitem{Secker2016}  T. Secker, R. Gerritsma, A. W. Glaetzle, and A. Negretti, \href{https://journals.aps.org/pra/abstract/10.1103/PhysRevA.94.013420} {Phys. Rev. A \textbf{94}, 013420  (2016)}.

\bibitem{Saito2017} R. Saito, S. Haze, M. Sasakawa, R. Nakai, M. Raoult, H. Da Silva, Jr., O. Dulieu, and T. Mukaiyama, \href{https://journals.aps.org/pra/abstract/10.1103/PhysRevA.95.032709} {Phys. Rev. A \textbf{95}, 032709 (2017)}.

\bibitem{Idziaszek2009}  Z. Idziaszek, T. Calarco, P. S. Julienne, and A. Simoni, \href{https://journals.aps.org/pra/abstract/10.1103/PhysRevA.79.010702} {Phys. Rev. A \textbf{79}, 010702  (2009)}.

\bibitem{Idziaszek2011}  Z. Idziaszek, A. Simoni, T. Calarco and P. S. Julienne,  \href{https://iopscience.iop.org/article/10.1088/1367-2630/13/8/083005} {New J. Phys. \textbf{13}, 083005  (2011)}.

\bibitem{Tomza2015}  M. Tomza, C. P. Koch, and R. Moszynski, \href{https://journals.aps.org/pra/abstract/10.1103/PhysRevA.91.042706} {Phys. Rev. A \textbf{91}, 042706  (2015)}.

\bibitem{Silva2015}  H. da Silva Jr, M. Raoult, M. Aymar, and O. Dulieu, \href{https://iopscience.iop.org/article/10.1088/1367-2630/17/4/045015} {New Journal of Physics \textbf{17}, 045015  (2015)}.

\bibitem{Belyaev2012} A. K. Belyaev, S. A. Yakovleva, M. Tacconi and F. A. Gianturco \href{https://link.aps.org/pdf/10.1103/PhysRevA.85.042716}{Phys. Rev. A. \textbf{85}, 042716 (2012)}.

\bibitem{Tacconi2011} M. Tacconi, F. A. Gianturco and A. K. Belyaev, \href{https://pubs.rsc.org/en/content/articlelanding/2011/cp/c1cp20916gstring#!divAbstract}{Phys. Chem. \textbf{13}, 19156 (2011)}.

\bibitem{Raab2009}  P. Raab and H. Friedrich, \href{https://journals.aps.org/pra/abstract/10.1103/PhysRevA.80.052705} {Phys. Rev. A \textbf{80}, 052705  (2009)}.

\bibitem{Inouye1998}  S. Inouye, M. R. Andrews, J. Stenger, H.-J. Miesner, D. M. Stamper-Kurn and W. Ketterle, \href{https://doi.org/10.1038/32354} {Nature \textbf{392}, 151–154  (1998)}.

\bibitem{Klein2017}  A. Klein, Y. Shagam, W. Skomorowski, P. S. Żuchowski, M. Pawlak, L. M. C. Janssen, N. Moiseyev, S. Y. T. van de Meerakker, A. van der Avoird, C. P. Koch and E. Narevicius  \href{https://doi.org/10.1038/nphys3904} {Nature Phys. \textbf{13}, 35–38 (2017)}.

\bibitem{Cetina2012} M. Cetina, A. T. Grier, and V. Vuletic, \href{https://journals.aps.org/prl/abstract/10.1103/PhysRevLett.109.253201} {Phys. Rev. Lett. \textbf{109}, 253201 (2012)}.

\bibitem{Meir2016} Z. Meir, T. Sikorsky, R. Ben-shlomi, N. Akerman, Y. Dallal and R. Ozeri, \href{https://journals.aps.org/prl/abstract/10.1103/PhysRevLett.117.243401} {Phys. Rev. Lett. \textbf{117}, 243401 (2016)}.

\bibitem{Pinkas2020} M. Pinkas, Z. Meir, T. Sikorsky, R. Ben-shlomi, N. Akerman and R. Ozeri, \href{https://iopscience.iop.org/article/10.1088/1367-2630/ab6792/pdf} {New J. Phys. \textbf{22}, 013047 (2020)}.

\bibitem{Feldker2020} T. Feldker, H. Fürst, H. Hirzler, N. V. Ewald, M. Mazzanti, D. Wiater, M. Tomza and R. Gerritsma, \href{https://doi.org/10.1038/s41567-019-0772-5} {Nat. Phys. \textbf{16}, 413–416 (2020)}.

\bibitem{Schmid2018} T. Schmid, C. Veit, N. Zuber, R. Löw, T. Pfau, M. Tarana, and M. Tomza \href{https://journals.aps.org/prl/abstract/10.1103/PhysRevLett.120.153401} {Phys. Rev. Lett. \textbf{120}, 153401 (2018)}.

\bibitem{Hall2013} F. H. J. Hall, P. Eberle, G. Hegi, M. Raoult, M. Aymar, O. Dulieu, and S. Willitsch, \href{https://www.tandfonline.com/doi/full/10.1080/00268976.2013.780107}{Mol. Phys. \textbf{111}, 2020 (2013)}. 

\bibitem{Hall2012}  F. H. J. Hall and S. Willitsch, \href{https://link.aps.org/doi/10.1103/PhysRevLett.109.233202}{Phys. Rev. Lett \textbf{109}, 233202  (2012)}.

\bibitem{Haze2013} S. Haze, S. Hata, M. Fujinaga, and T. Mukaiyama, \href{https://journals.aps.org/pra/pdf/10.1103/PhysRevA.87.052715} {Phys. Rev. A. \textbf{87}, 052715 (2013)}.

\bibitem{Dorfler2019}  A.D. Dörfler, P. Eberle, D.  Koner, M. Tomza, M. Meuwly and S. Willitsch,  \href{https://www.nature.com/articles/s41467-019-13218-x.pdf} {Nat. Commun \textbf{10}, 5429 (2019)}.

\bibitem{Zipkes2010}  C. Zipkes, S. Palzer, and M. Köhl, \href{https://www.nature.com/articles/nature08865}{Nature \textbf{464}, 388 (2010)}.

\bibitem{Schmid2010} S. Schmid, A. Härter, and J. H. Denschlag, \href{https://link.aps.org/doi/10.1103/PhysRevLett.105.133202}{Phys. Rev. Lett. \textbf{105}, 133202 (2010)}.

\bibitem{Schmid2012}  S. Schmid, A. Härter, A. Frisch, S. Honika and J.H. Denschlag,  \href{https://aip.scitation.org/doi/pdf/10.1063/1.4718356?class=pdf} {Rev. Sci. Instrum \textbf{83}, 053108 (2012)}.

\bibitem{Bell2009}  M. T. Bell, A. D. Gingell, J. Oldham, T. P. Softley, S. Willitsch, \href{https://www.ncbi.nlm.nih.gov/pubmed/20151539}, Faraday Discuss. \textbf{142}, 73 (2009).

\bibitem{Grier2009}  A. T. Grier, M. Cetina, F. Oručević, and V. Vuletić,  \href{https://journals.aps.org/prl/pdf/10.1103/PhysRevLett.102.223201} {Phys. Rev. Lett. \textbf{102}, 223201 (2009)}.

\bibitem{Puri2019}  P. Puri, M. Mills, I. Simbotin, J. A. Montgomery Jr, R. Côté, C. Schneider, A. G. Suits and E. R. Hudson,  \href{https://www.nature.com/articles/s41557-019-0264-3?draft=collection} {Nature Chemistry \textbf{11}, 615–621 (2019)}.

\bibitem{Eberle2016}  P. Eberle, A. D. Dörfler  C. von Planta, K. Ravi and S. Willitsch,  \href{https://onlinelibrary.wiley.com/doi/full/10.1002/cphc.201600643} {ChemPhysChem \textbf{17}, 3769–3775 (2016)}.

\bibitem{Puri2018}  P. Puri, M. Mills, E. P. West, C. Schneider and E. R. Hudson,  \href{https://aip.scitation.org/doi/10.1063/1.5031145} {Rev. Sci. Instrum \textbf{89}, 083112 (2018)}.

\bibitem{Ruti2019} R. Ben-Shlomi, R. Vexiau, Z. Meir, T. Sikorsky, N. Akerman, M. Pinkas, O. Dulieu, and R. Ozeri, \href{http://arxiv.org/abs/1907.06736}{arXiv, \textbf{1907.06736} (2019)}.

\bibitem{Hall2011} F. H. J. Hall, M. Aymar, N. Bouloufa-Maafa, O. Dulieu, and S. Willitsch,\href{https://link.aps.org/doi/10.1103/PhysRevLett.107.243202}{Phys. Rev. Lett. \textbf{107}, 243202 (2011)}.

\bibitem{Meir2018} Z. Meir, T. Sikorsky, R. Ben-shlomi, N. Akerman, M. Pinkas, Y. Dallal and R. Ozeri, \href{https://www.tandfonline.com/doi/full/10.1080/09500340.2017.1397217} {Journal of Modern Optics \textbf{65}, 501-519 (2018)}.

\bibitem{Schmid2006}  S. Schmid, G. Thalhammer, K. Winkler, F. Lang, J.H. Denschlag,  \href{https://iopscience.iop.org/article/10.1088/1367-2630/8/8/159} {New Journal of Physics \textbf{8}, 1–75 (2006)}.

\bibitem{Langevin1905} P. Langevin, Ann. Chim. Phys. \textbf{5}, 245 (1905).

\bibitem{singleshot} Z. Meir, T. Sikorsky, N. Akerman, R. Ben-shlomi, M. Pinkas and R. Ozeri, \href{https://link.aps.org/doi/10.1103/PhysRevA.96.020701}{Phys. Rev. A. \textbf{96}, 020701 (2017)}.

\bibitem{Zipkes2011}  C. Zipkes, L. Ratschbacher, C. Sias and M. Kohl,  \href{http://iopscience.iop.org/1367-2630/13/5/053020} {New Journal of Physics \textbf{13}, 053020 (2011)}.

\bibitem{DeVoe2009}  R. G. DeVoe, \href{https://link.aps.org/doi/10.1103/PhysRevLett.102.063001} {Phys. Rev. Lett \textbf{102}, 063001 (2009)}.


\bibitem{Chen2014}  K. Chen, S. T. Sullivan, and E. R. Hudson, \href{https://journals.aps.org/prl/abstract/10.1103/PhysRevLett.112.143009} {Phys. Rev. Lett.
\textbf{112}, 143009 (2014)}.

\bibitem{Rouse2017} I. Rouse and S. Willitsch, \href{https://journals.aps.org/prl/abstract/10.1103/PhysRevLett.118.143401} {Phys. Rev. Lett. \textbf{118}, 143401 (2017)}.

\bibitem{Demortier2007} L. Demortier, \href{https://www-cdf.fnal.gov/~luc/statistics/cdf8662.pdf}{CDF/MEMO/STATISTICS/PUBLIC/8662, The Collider Detector at Fermilab notes (2007)}.
\end{references}
\end{document}